\documentclass[11pt]{amsart}
\usepackage[utf8]{inputenc}
\usepackage[T1]{fontenc}
\usepackage[francais,english]{babel}
\usepackage{amssymb}
\usepackage{mathtools}
\usepackage{color}

\everymath{\displaystyle}
\usepackage{geometry}
\usepackage{subcaption}
\usepackage{dsfont}
\usepackage{epsfig}
\usepackage{colortbl}

\usepackage{hyperref}

\def\@xthm#1#2{\@beginassumption{#2}{\csname the#1\endcsname}{}\ignorespaces}
\def\@ythm#1#2[#3]{\@opargbeginassumption{#2}{\csname the#1\endcsname}{#3}\ignorespaces}%
\def\@beginassumption#1#2#3{\par\addvspace{8pt plus3pt minus2pt}%
              \noindent{\csname#1headfont\endcsname#1\ \ignorespaces#3 #2.}%
              \csname#1font\endcsname\hskip.5em\ignorespaces}
\def\@endassumption{\par\addvspace{8pt plus3pt minus2pt}\@endparenv}
\input epsf
\begin{document}

%........................................................................
\input epsf
%definitions
\def\RR{\mathbb{R}}
\def\NN{\mathbb{N}}
\def\SS{\mathbb{S}}

\def\cG{{\cal G}}
\def\cB{{\cal B}}
\def\cC{{\cal C}}
\def\cD{{\cal D}}

\def\cI{{\cal I}}
\def\cL{{\cal L}}
\def\cJ{{\cal J}}
\def\cA{{\cal A}}
\def\cS{{\cal S}}
\def\cP{{\cal P}}
\def\cH{{\cal H}}

\def\cF{\mathcal{F}}

\newcommand{\bnu}{\boldsymbol{\nu}}
\newcommand{\bomega}{\boldsymbol{\omega}}
\newcommand{\btau}{\boldsymbol{\tau}}

\newcommand{\bx}{{\boldsymbol{x}}}
\newcommand{\by}{{\boldsymbol{y}}}
\newcommand{\bbj}{{\boldsymbol{j}}}
\newcommand{\bn}{{\boldsymbol{n}}}
\newcommand{\bq}{{\boldsymbol{q}}}
\newcommand{\bbi}{{\boldsymbol{i}}}
\newcommand{\bv}{{\boldsymbol{v}}}
\newcommand{\bu}{{\boldsymbol{u}}}
\newcommand{\bw}{{\boldsymbol{w}}}
\newcommand{\bz}{{\boldsymbol{z}}}

\newcommand{\bbr}{\mathbb R}
\newcommand{\bbc}{\mathbb C}
\newcommand{\bbs}{\mathbb S}
\newcommand{\mE}{\mathbb E}
\newcommand{\bbp}{\mathbb P}
\newcommand{\bbt}{\mathbb T}
\newcommand{\bbm}{\mathbb M}
\newcommand{\bbn}{\mathbb N}

\def\b{{\beta}}
\def\a{{\alpha}}
\def\d{{\delta}}
\def\g{{\gamma}}
\def\vf{{\varphi}}
\def\ve{{\varepsilon}}
\def\l{{\lambda}}
\def\th{{\theta}}

\def\p{\partial}

\def\div{\mbox{{\rm div}}}
\def\Div{\mbox{{\rm Div}}}

\def\no{\noindent}
\def\fo{{\forall \,}}
\def\sm{\smallskip}
\def\me{\medskip}
\def\bi{\bigskip}

\def\vb{,}
\def\pb{\, .}
\def\pb{.}
\def\pa{\cdotp}
\def\va{\raise 2pt\hbox{,}}

\font\dc=cmbxti10 %slanted boldface for titles of "definition"

\overfullrule=0pt

\def\bull{{\vrule height.9ex width.8ex depth.1ex}}

\def\boxitt#1{\vbox{\hrule\hbox{\vrule\kern3pt
\vbox{\kern7pt#1\kern7pt}\kern3pt\vrule}\hrule}}

\markboth{N.~Bellomo, Authors}{Towards a Mathematical Theory  of Behavioral Swarms}

%%%%%%%%%%%%%%%%%%% Publisher's Area please ignore %%%%%%%%%%%%%%%%%%%%%%%
%
%\catchline{}{}{}{}{}
%
%%%%%%%%%%%%%%%%%%%%%%%%%%%%%%%%%%%%%%%%%%%%%%%%%%%%%%%%%%%%%%%%%%%%%%%%%%

\title{Towards a Mathematical Theory  of Behavioral Swarms}

\author{N.~Bellomo$^{(1)}$, S.-Y.~Ha$^{(2)}$, N.~Outada$^{(3)}$}

\address{$^{(1)}$University of Granada, Departamento de Matem\'atica Aplicada, Spain \\ and Politecnico of Torino, Italy.  \\nicola.bellomo@ugr.es, nicola.bellomo@polito.it}

\address{$^{(2)}$Seoul National University, Department of Mathematical Sciences and Research Institute of Mathematics\\
Seoul National University, Seoul 08826 and \\
Korea Institute for Advanced Study, Hoegiro, 85, Seoul 02455, Republic of Korea\\
syha@snu.ac.kr}

\address{$^{(3)}$ Cadi Ayyad University, Faculty of Sciences Semlalia, LMDP, Morocco and UMMISCO (IRD-Sorbonne University, France)\\
outada@ljll.math.upmc.fr}

% \vspace{1cm} March 25, 2018}

%\footnote{This author has been funded with support
%from the European Commission. This publication reflects the views
%only of the author, and the Commission cannot be held responsible
%for any use which may be made of the information contained
%therein.}

\maketitle

%\begin{history}
%\received{(Day Month Year)} \revised{(Day Month Year)}
%%\accepted{(Day Month Year)}
%\comby{(xxxxxxxxxx)}
%\end{history}

\begin{abstract}
This paper presents a unified mathematical theory of swarms where the dynamics of social behaviors interacts with the mechanical dynamics of self-propelled particles. The term \textit{behavioral swarms} is introduced to characterize the specific object of the theory which is subsequently followed by applications. As concrete examples for our unified approach, we show that several Cucker-Smale type models with internal variables fall down to our framework. Subsequently the modeling goes beyond the  Cucker-Smale approach and  looks ahead to research perspectives.
\end{abstract}
\vspace{0.25cm}

\keywords{{\em Keywords:} Collective dynamics, Cucker-Smale flocking, learning, living complex systems, self-organization, swarming, collective behavior, nonlinear interactions}
\vspace{0.25cm}

\subjclass{AMS Subject Classification: 82D99, 91D10}

\parindent=20pt

%%%%%%%%%%%%%%%%%%%%%%%%%%%%%%%%%%%%%%%%
%%%%%%%%%%%%%%%%%%%%%%%%%%%%%%%%%%%%%%%%
\section{Plan of the paper}\label{sec:1}
%%%%%%%%%%%%%%%%%%%%%%%%%%%%%%%%%%%%%%%%
%%%%%%%%%%%%%%%%%%%%%%%%%%%%%%%%%%%%%%%%

The celebrated paper by Cucker and Smale~\cite{[CS07]} has boosted a huge literature on the modeling, qualitative analysis, and computational applications of the mathematical theory of swarms. Namely, models which describe, within  a pseudo-Newtonian framework, the mechanics of many interacting self-propelled particles. Interactions are generally nonlocal and nonlinearly additive. These produce accelerations, where inertia is hidden in the interaction parameters rather than being explicitly taken into account. The mathematical literature in the field has been reviewed and critically analyzed in~\cite{[ALBI19]}, specifically treated in Sections 5 and 6 which are devoted to the theory of swarms.  This topic has been  related to the dynamics of crowds as well as to computational models to treat kinetic and multi-agent systems.

After Cucker-Smale's (C-S)'s seminal work, the (C-S) model has been extensively studied in literature, for example, we refer to \cite{[CHL17]} and references therein for mathematical aspects of the model. Among them, we briefly discuss some recent works incorporating social behaviors together with mechanical aspect of the model. Original C-S model describes the temporal evolution of the mechanical variables (positions and momentum of C-S particles), whereas in the modeling for the collective dynamics of biological and social complex systems, one needs to take into account of internal variables such as temperature, spin and excitation \cite{[HKKS19],[HL09],[HR17]} etc, see Section \ref{sect:3}.

Recently, in \cite{[HKPZ19]}, the authors found that the one-dimensional C-S model can be converted into the first-order nonlinear consensus model with monotone coupling function on the real line. Then, in the analogy with the corresponding result for the Kuramoto model, they showed that for a given initial data and coupling strength, they provided an algorithm determining the number of asymptotic clusters and their group velocities which can be called "the complete cluster predictability" of the one-dimensional C-S model. Unfortunately, there is no multi-dimensional counterpart for this complete cluster predictability. The clustering dynamics of the C-S model under the attractive-repulsive couplings was also discussed in \cite{[FHJ19],[HJKK12],[KHKJ14]}

Swarms theory often interacts with the kinetic theory approaches which might be based on Fokker-Plank methods~\cite{[PT13]} or stochastic evolutive games within the framework of the so-called kinetic theory of active particles~\cite{[BBGO17]}. An example of a kinetic theory approach to swarm modeling is given in~\cite{[BH17]}.

A recent literature has shown a growing interest on the applications of swarm theory to model social and economical problems~\cite{[ABH13],[BCLY17],[BCK19],[BDKMT19]}, while the literature swarm on predator-preys system~\cite{[BS12]} suggests to introduce behavioral dynamics. Analogous requirements are induced in the control problems of sparse agents\cite{[APTZ17],[AP18],[PPT18]}.

Our paper accounts for all aforementioned hints and proposes a development of a mathematical theory of swarms where the dynamics of social behaviors, including  emotional states, interacts with the mechanical dynamics of self-propelled particles, where the term \textit{behavioral swarms} is introduced to denote the specific class of dynamical systems under consideration. The theory proposed in this paper is subsequently followed by applications. In more detail, the contents is delivered as follows.

Section 2 presents the main results of our paper, namely the aforementioned theory, where a general framework includes the interaction between mechanics and social behaviors, hence towards a theory of behavioral swarms. The theory refers to both second order  and hybrid first-second order models. It leads to  general frameworks deemed to provide the conceptual basis for the derivation of specific models.

Section 3  shows how  C-S models with internal variables can be referred the general structure derived in Section 2 accounting for the heterogeneity features accounted for by our approach. Models refer to the hybrid structures, where the dynamics of the social variable is described by first order equations, while accelerations are included for the mechanical variable.

 Section 4 shows, by selecting a number of simple case studied how  models which go beyond the C-S approach can be derived. The derivation  refer to the mathematical structures which include accelerations for all dependent mechanical variables accounting for the role of the social dynamics over the motion by a first order model. Some simulations presented to enlighten how the activity variables modifies the trajectories of the motion.

 Section5 presents  some research perspectives which are brought to the attention of the reader looking ahead to different fields of applications, for instance biology.

%%%%%%%%%%%%%%%%%%%%%%%%%%%%%%%%%%%%%%%%%%%%%%%%%%%%%%%%%%%%%%%%%%%%%%%%%%
%%%%%%%%%%%%%%%%%%%%%%%%%%%%%%%%%%%%%%%%%%%%%%%%%%%%%%%%%%%%%%%%%%%%%%%%%%
\section{Towards a mathematical theory of behavioral swarms}\label{sect:2}
%%%%%%%%%%%%%%%%%%%%%%%%%%%%%%%%%%%%%%%%%%%%%%%%%%%%%%%%%%%%%%%%%%%%%%%%%%
%%%%%%%%%%%%%%%%%%%%%%%%%%%%%%%%%%%%%%%%%%%%%%%%%%%%%%%%%%%%%%%%%%%%%%%%%%

Let us consider  the dynamics of an heterogeneous  swarm  of $N$ interacting self-propelled particles, labeled by the subscript $i=1,\ldots,N$, where heterogeneity is also induced by the ability of each particle to express a scalar social state which will be called \textit{activity}. The individual state of the $i$-th particle can be defined by position $\bx_i$, velocity $\bv_i$ and activity $u_i$, while the speed of the activity variable, namely the time derivative of $u_i$ is denoted by $z_i$. We firstly consider the dynamics of one population only, subsequently the generalization to several interacting populations
is treated.

The velocity  $\bv_i$  can be represented in polar coordinates as follows:
\begin{equation}
\bv_i = \{v_i, \bomega_i\}, \hskip.5cm  v_i \in[0,1],  \hskip.5cm \bomega_i = \frac{\bv_i}{||\bv_i||}\va
\end{equation}
where the unit vector $\bomega_i$ can be represented by the angular variables referred to a  cartesian system and where the cartesian components of the  position $\bx_i$ are referred to a characteristic length $\ell$ of the system, while $v_i$ has been referred to the limit velocity $v_M$ which can be reached by the fastest particle. In particular, if the system is localized in a bounded domain $\Sigma$ the $\ell$ is the diameter of the circle containing $\Sigma$, while if the system moves  in an unbounded domain, the $\ell$ is simply referred to the domain  $\Sigma_0$  containing the particles at $t=0$.

Since the self-propelled particles express a social function, the term \textit{active particle} or, in short, \textit{a-particle} will be used to denote them, while the individual a-particle is denoted by \textit{i-particle}. The state of each i-particle is called \textit{micro-state}, while the set of all positions, velocities, and activities of the whole system is denoted by $\bx$, $\bv$ and $\bu$.

In general, dimensionless variables are used dividing the linear component of the position by a characteristic length $\ell$ related to the geometry of the system, and the speed by the highest speed reachable by the specific active particles object of the modeling approach. The objective consists in  referring all macroscopic states, namely geometrical, mechanical, and social, to the a domain of the order of unity. Accordingly, the activity variable $u_i$ takes values in $[0,1]$, where the limit values $u_i=0$ and $u_i=1$ represent, respectively, the lowest and highest values of the activity.

 Bearing all the above in mind let us consider the derivation of a general mathematical structure suitable to provide the conceptual framework for the derivation of models. This new  conceptual approach should account, at least, for the following features:
\begin{enumerate}
\item Each particle is able to develop a specific strategy which is heterogeneously distributed.

\vskip.1cm \item Interactions can be nonlocal and  nonlinearly additive.

\vskip.1cm \item A decisional hierarchy can be used by assuming that interactions first modify the activity and subsequently the motion which depends also on the activity.

\vskip.1cm \item The approach is such that firstly mathematical structures are derived  to provide the conceptual framework for the derivation of models, and subsequently this structure is implemented by models  interactions which lead to specific models.

\vskip.1cm \item  Each a-particle has sensitivity domain and interacts with all particles within the said domain.
\end{enumerate}

The following quantities can be introduced  to model, still at a formal level, interactions:

\vskip.1cm \noindent $\eta_i$ models the interaction rate of individual based interactions between $i$-particle with all particles in the sensitivity domain.

\vskip.1cm \noindent $\Omega_i$ is the vision-based sensitivity domain of the $i$-particle.

\vskip.1cm \noindent $\vf_i$ denotes the action, which occurs with rate  $\eta_i$,  over the activity variable over the $i$-particle by all particles in $\Omega_i$.

\vskip.1cm \noindent $\psi_i$ denotes the acceleration, which occurs with rate  $\eta_i$,  over the mechanical  variable by all particles in $\Omega_i$.

\vskip.2cm \noindent $\bullet$ \textit{Second order mechanics and activity:}
In general, all above quantities depend on all variables $\bw = \bx,\bv, \bu, \bz$ to be selected within the sensitivity domain. The formal structure of the framework is as follows:
\begin{equation}\label{swarm-s}
\begin{cases}
\displaystyle  \frac{d\bx_{i}}{dt} =  \bv_i, \\[3mm]
\displaystyle \frac{d\bv_{i}}{d t} =  \sum_{j \in \Omega_i} \eta_i(\bx_i, \bv_i, \bu_i, \bx_j,\bv_j, \bu_j) \, \psi_i(\bx_i, \bv_i, \bu_i, \bx_j,\bv_j, \bu_j),\\[3mm]
\displaystyle \frac{d u_i}{dt} =  z_i, \\[3mm]
\displaystyle \frac{d z_i}{d t} = \sum_{j \in \Omega_i} \eta_i(\bx_i, \bv_i, \bu_i, \bx_j,\bv_j, \bu_j) \, \vf_i(\bx_i, \bv_i, \bu_i, \bx_j,\bv_j, \bu_j),
 \end{cases}
\end{equation}
where the notation $j \in \Omega_i$ indicates that summation refers to all $j$-particles in the domain $\Omega_i$.

The structure (\ref{swarm-s}) consists in  a second order framework as it involves acceleration terms for all variables, first order frameworks can be obtained by simplified models leading to velocity, rather than acceleration, for both activity and mechanical variables. However, the modeling approach can consider hybrid frameworks which are first order for the activity variable and second order for the mechanical variables. An example is reported, at a  formal level, in the following, where the dynamics of the activity variable is of first order, while that of the mechanical variable is classically second order.
\newpage

\vskip.2cm \noindent $\bullet$ \textit{Second order mechanics and first order activity:}
\begin{equation}\label{swarm-h}
\begin{cases}
\displaystyle  \frac{d\bx_{i}}{dt} =  \bv_i, \\[3mm]
\displaystyle \frac{d\bv_{i}}{d t} =  \sum_{j \in \Omega_i} \eta_i(\bx_i, \bv_i, \bu_i, \bx_j,\bv_j, \bu_j) \, \psi_i(\bx_i, \bv_i, \bu_i, \bx_j,\bv_j, \bu_j),\\[3mm]
\displaystyle \frac{d u_i}{d t} = \sum_{j \in \Omega_i} \eta_i(\bx_i, \bv_i, \bu_i, \bx_j,\bv_j, \bu_j) \, \mu_i(\bx_i, \bv_i, \bu_i, \bx_j,\bv_j, \bu_j),
 \end{cases}
\end{equation}
with obvious meaning of the terms $\mu_i$ and where all components of the dependent variables, which play a role in the dynamics, have been explicitly indicated. By analogous calculations we can consider a formal structure for first order models both on the mechanical and the activity variable, however we do not consider this specific case as it appears to be too artificial with respect  to rules of mechanics.

\vskip.2cm \noindent $\bullet$ \textit{Second order mechanics and  activity for a mixture of functional subsystems:}
It can be shown how the formal structures (\ref{swarm-s}) and  (\ref{swarm-h}) can be generalized to study the dynamics of a-particles grouped into $n$  populations which can be called \textit{functional subsystems}, in short FSs, labeled by the subscripts $k = 1, \ldots, n$. Formal calculations and obvious generalization of notations yield:
\begin{equation}\label{multi-swarm-s}
\begin{cases}
\displaystyle  \frac{d\bx_{ik}}{dt} =  \bv_{ik}, \\[3mm]
\displaystyle \frac{d\bv_{ik}}{d t} = \sum_{{jq} \in \Omega_{ik}} \eta_{ik}(\bx_{ik}, \bv_{ik}, \bu_{ik}, \bz_{ik}, \bx_{jq}, \bv_{jq}, \bu_{jq}, \bz_{jq})  \\[2mm]
\hskip3cm \times\, \psi_{ik}(\bx_{ik}, \bv_{ik}, \bu_{ik}, \bz_{ik}, \bx_{jq}, \bv_{jq}, \bu_{jq}, \bz_{jq}), \\[3mm]
\displaystyle \frac{d u_{ik}}{dt} =  z_{ik}, \\[3mm]
\displaystyle \frac{d z_{ik}}{d t} =  \sum_{{jq} \in \Omega_{ik}} \eta_{ik}(\bx_{ik}, \bv_{ik}, \bu_{ik}, \bz_{ik}, \bx_{jq}, \bv_{jq}, \bu_{jq}, \bz_{jq})  \\[2mm]
\hskip3cm \times\, \vf_{ik}(\bx_{ik}, \bv_{ik}, \bu_{ik}, \bz_{ik}, \bx_{jq}, \bv_{jq}, \bu_{jq}, \bz_{jq}).
  \end{cases}
\end{equation}
The mathematical structure \eqref{multi-swarm-s} can be simplified under the hypothesis that the sensibility  domain of the $ik$-particle contains only the a-particles which belong to the same functional subsystem as the $ik$-particle, namely the $k$-th functional subsystem. The mathematical structure can be
particularized in this case as follows:
\newpage
\begin{equation}\label{multi-swarm-s-1}
\begin{cases}
\displaystyle  \frac{d\bx_{ik}}{dt} =  \bv_{ik}, \\[3mm]
\displaystyle \frac{d\bv_{ik}}{d t} = \sum_{{j} \in \Omega_{ik}} \eta_{ik}(\bx_{ik}, \bv_{ik}, \bu_{ik}, \bz_{ik}, \bx_{jk}, \bv_{jk}, \bu_{jk}, \bz_{jk})  \\[2mm]
\hskip3cm \times\, \psi_{ik}(\bx_{ik}, \bv_{ik}, \bu_{ik}, \bz_{ik}, \bx_{jk}, \bv_{jk}, \bu_{jk}, \bz_{jk}), \\[3mm]
\displaystyle \frac{d u_{ik}}{dt} =  z_{ik}, \\[3mm]
\displaystyle \frac{d z_{ik}}{d t} =  \sum_{{j} \in \Omega_{ik}} \eta_{ik}(\bx_{ik}, \bv_{ik}, \bu_{ik}, \bz_{ik}, \bx_{jk}, \bv_{jk}, \bu_{jk}, \bz_{jk})  \\[2mm]
\hskip3cm \times\, \vf_{ik}(\bx_{ik}, \bv_{ik}, \bu_{ik}, \bz_{ik}, \bx_{jk}, \bv_{jk}, \bu_{jk}, \bz_{jk}).
  \end{cases}
\end{equation}

Analogous calculations yield:
\vskip.2cm \noindent $\bullet$ \textit{Second order mechanics and first order activity for a mixture of functional subsystems:}
\begin{equation}\label{multi-swarm-h}
\begin{cases}
\displaystyle  \frac{d\bx_{ik}}{dt} =  \bv_{ik}, \\[3mm]
\displaystyle \frac{d\bv_{ik}}{d t} = \sum_{{jq} \in \Omega_{ik}} \eta_{ik}(\bx_{ik}, \bv_{ik}, \bu_{ik}, \bx_{jq}, \bv_{jq}, \bu_{jq}) \\[2mm]
\hskip3cm \times\, \psi_{ik}(\bx_{ik}, \bv_{ik}, \bu_{ik}, \bx_{jq}, \bv_{jq}, \bu_{jq}), \\[3mm]
\displaystyle \frac{d u_{ik}}{dt} = \sum_{{jq} \in \Omega_{ik}} \eta_{ik}(\bx_{ik}, \bv_{ik}, \bu_{ik}, \bx_{jq}, \bv_{jq}, \bu_{jq}) \\[2mm]
\hskip3cm \times \, \mu_{ik}(\bx_{ik}, \bv_{ik}, \bu_{ik}, \bx_{jq}, \bv_{jq}, \bu_{jq}),
  \end{cases}
\end{equation}
where all components of the dependent variables, which play a role in the dynamics, appear in the argument of the interaction terms.
\vskip.2cm

A detailed characterization of these mathematical structures, namely of the interaction terms $ \Omega_{ik}, \eta_{ik},  \vf_{ik}$ and $\psi_{ik}$, can be developed only for well defined case studies. However, some common features of the modeling approach can be given by a qualitative description to be formalized for each case study, as we shall see in the next sections.
\begin{itemize}
\item  \textbf{Hierarchy:} The study of human crowds~\cite{[BGO19]} has suggested a hierarchy in the decision making by which walkers develop their walking strategies, namely by interactions which firstly induce modification of the social state, subsequently walkers modify their walking direction, and finally they adapt the speed to the local new flow direction. Introducing a hierarchy is an important step of the modeling approach. Arguably, an analogous sequence of dynamical choice can be developed for a broad variety of case studies.

\vskip.1cm \item  \textbf{Sensitivity domain:} This quantity is defined by a cone with vertex in $\bx_i$, vertex angle $\Theta$, and  with  axis along the velocity. $\Omega_i$ is finite being  truncated at a distance $R$ which might be related to the critical finite number of active particles which have a sensitive influence.  The sensitivity domain might be modified by visibility problem.

\vskip.1cm \item  \textbf{Interaction rate:} The modeling of the interaction rates can be referred to a distance between the interacting entities by a metric suitable to account both for the distance between the interacting entities and that of their statistical distribution.

\vskip.1cm \item  \textbf{Social action:} The social action depends on the interaction of each particle with those in its sensitivity domain. It is specific of each system to be modeled.

\vskip.1cm \item  \textbf{Mechanical action:} Mechanical actions follow the rules of classical mechanics, but the parameters leading to accelerations depend on on the social state by models to be properly defined.
\end{itemize}

%%%%%%%%%%%%%%%%%%%%%%%%%%%%%%%%%%%%%%%%%%%%%%%%%%%%%%%%%%%%%%%%%%%%%%%%
%%%%%%%%%%%%%%%%%%%%%%%%%%%%%%%%%%%%%%%%%%%%%%%%%%%%%%%%%%%%%%%%%%%%%%%%
\section{Cucker-Smale type models with internal variables}\label{sect:3}
%%%%%%%%%%%%%%%%%%%%%%%%%%%%%%%%%%%%%%%%%%%%%%%%%%%%%%%%%%%%%%%%%%%%%%%%
%%%%%%%%%%%%%%%%%%%%%%%%%%%%%%%%%%%%%%%%%%%%%%%%%%%%%%%%%%%%%%%%%%%%%%%%

In this section, we provide several C-S type models which fit into our proposed general framework~\eqref{swarm-h}. Our paper, as mentioned, aims at going beyond this pioneer model, hence this section can be viewed as a preliminary step towards further developments.

Bearing in mind this introductory remark,  we  first present the original model for \textit{mechanical particles} and, subsequently, we introduce some Cucker-Smale type models with internal variables related to~\eqref{swarm-h}.

%%%%%%%%%%%%%%%%%%%%%%%%%%%%%%%%%%%%%%%%%%%%%%%%%%%%
\subsection{The Cucker-Smale model}\label{sect:3.1}
%%%%%%%%%%%%%%%%%%%%%%%%%%%%%%%%%%%%%%%%%%%%%%%%%%%%

Consider an ensemble of mechanical Cucker-Smale particles whose states are represented by position and velocity variables. More precisely, let $\bx_i$ and $\bv_i$ be the position and velocity of the $i$-th C-S particle in the free Euclidean space $\bbr^d$, respectively. Then, the temporal evolution of mechanical variables $(\bx_i, \bv_i)$ is governed by the Newton-like system:
\begin{equation} \label{C-0}
\begin{cases}
\displaystyle \frac{d{\bx}_i}{dt} = \bv_i, \quad t > 0, \quad 1 \le i \le N,\\[3mm]
\displaystyle \frac{d\bv_i}{dt} = \frac{\kappa}{N}\sum_{j=1}^N \phi(\|\bx_j - \bx_i\|) \left(\bv_j-\bv_i \right),
\end{cases}
\end{equation}
where $\|\cdot\|$ denotes the standard $\ell^2$-norm in $\bbr^d$, $\kappa$ represents the strength of coupling and the communication weight $\phi: [0,\infty) \to \bbr_+$ is bounded, Lipschitz continuous and monotonically decreasing:
\begin{align}
\begin{aligned} \label{C-1}
& 0 < \phi(r) \le \phi(0), \quad [\phi]_{Lip} := \sup_{x \not = y} \frac{|\phi(x) - \phi(y)|}{|x-y|} < \infty, \\[3mm]
&  (\phi(r)-\phi(s))(r-s) \le 0, \quad r,s\geq 0.
\end{aligned}
\end{align}

System \eqref{C-0} does not conserve the speed of particles. Thus, as a slight variation of system \eqref{C-0}, we introduce the Cucker-Smale model with unit speed. For this, we replace the velocity coupling term $\bv_j-\bv_i$ by $\bv_j - (\bv_j \cdot \bv_i) \bv_i$ which results in
\begin{equation} \label{C-1-1}
\begin{cases}
\displaystyle \frac{d{\bx}_i}{dt} = \bv_i, \quad t > 0, \quad 1 \le i \le N,\\[3mm]
\displaystyle \frac{d\bv_i}{dt} = \frac{\kappa}{N}\sum_{j=1}^N \phi(\|\bx_j - \bx_i\|) \left( \bv_j - (\bv_j \cdot \bv_i) \bv_i \right),
\end{cases}
\end{equation}
Then, it is easy to see that system \eqref{C-1-1} preserves the speed of particles:
\[ \|\bv_i(t) \| = \|\bv_i^0 \|, \quad t > 0, \quad i = 1, \cdots, N. \]
In the following four subsections, we provide four active C-S particles with internal variables such as excitation level, temperature, spin and phase which takes the form:
\begin{equation}\label{C-2}
\begin{cases}
\displaystyle  \frac{d\bx_{i}}{dt} =  \bv_i, \quad i = 1, \cdots, N, \\[3mm]
\displaystyle \frac{d\bv_{i}}{d t} =  \frac{\kappa_1}{N} \sum_{j=1}^{N} \psi_1( \bx_i, \bx_j) C_1(\bv_i, \bv_j, u_i, u_j), \\[3mm]
\displaystyle \frac{d u_i}{dt} =  \frac{\kappa_2}{N}  \sum_{j=1}^{N}  \psi_2( \bx_i, \bx_j) C_2(\bv_i, \bv_j, u_i, u_j).
\end{cases}
\end{equation}
Note that the choice
\[  \kappa_2 = 0, \quad  \kappa_1=\kappa, \quad \psi_1( \bx_i, \bx_j)  = \phi(\|\bv_j - \bv_i \|), \quad C_1(\bv_i, \bv_j, u_i, u_j) = \bv_j - \bv_i  \]
corresponds to the C-S model \eqref{C-0}.

%%%%%%%%%%%%%%%%%%%%%%%%%%%%%%%%%%%%%%%%%%%%%%%%%%%%%%%%%%%%%%%%%%
\subsection{Thermodynamic Cucker-Smale model} \label{sect:3.2}
%%%%%%%%%%%%%%%%%%%%%%%%%%%%%%%%%%%%%%%%%%%%%%%%%%%%%%%%%%%%%%%%%%
In this subsection, we introduce the Cucker-Smale model \cite{[HKR18],[HR17]} for thermodynamic particles. Consider an ensemble of C-S particles with temperatures as an internal variable. For this, we denote $u_i \in \bbr$ to be the temperature of the $i$-th C-S particle. Then, the thermodynamic Cucker-Smale model reads as follows.
\begin{equation}\label{C-4}
\begin{cases}
\displaystyle \frac{d\bx_i}{dt} = \bv_i, \\[3mm]
\displaystyle \frac{d\bv_i}{dt} = \frac{\kappa_1}{N} \sum_{j=1}^{N} \phi_1(\| \bx_j - \bx_i \|) \Big( \frac{\bv_j - \bv_c}{u_j} -  \frac{\bv_i - \bv_c}{u_i} \Big), \\[3mm]
\displaystyle \frac{d}{dt}\left(u_i + {{\frac{1}{2} \| \bv_i \|^2}}\right) = \frac{\kappa_2}{N}  \sum_{j=1}^{N} \phi_2(\| \bx_j - \bx_i \|)
\Big( \frac{1}{u_i} - \frac{1}{u_j} \Big)  \\[3mm]
\displaystyle \hspace{3cm} +  \frac{\kappa_1}{N} \sum_{j=1}^{N}  \phi_1(\| \bx_j - \bx_i \|)
\Big( \frac{\bv_j - \bv_c}{u_j} -  \frac{\bv_i - \bv_c}{u_i} \Big) \cdot \bv_c,
\end{cases}
\end{equation}
where $\bv_c := \frac{1}{N} \sum_{i=1}^{N} \bv_i$, and  $\kappa_1$ and $\kappa_2$ denote positive coupling strengths. \newline

Note that system \eqref{C-4} is Galilean invariant, so without loss of generality, we can choose the center of mass as origin of reference frame $\bv_c = 0.$ Thus, system \eqref{C-4} becomes
\begin{equation}\label{C-4-1}
\begin{cases}
\displaystyle \frac{d\bx_i}{dt} = \bv_i,  \\[3mm]
\displaystyle \frac{d\bv_i}{dt} = \frac{\kappa_1}{N} \sum_{j=1}^{N} \phi_1(\| \bx_j - \bx_i \|) \Big( \frac{\bv_j}{u_j} -  \frac{\bv_i}{u_i} \Big), \\[3mm]
\displaystyle \frac{d}{dt}\left(u_i + {{\frac{1}{2} \| \bv_i \|^2}}\right) = \frac{\kappa_2}{N}  \sum_{j=1}^{N} \phi_2(\| \bx_j - \bx_i \|)
\Big( \frac{1}{u_i} - \frac{1}{u_j} \Big).
\end{cases}
\end{equation}
The model \eqref{C-4} takes into account the mutual interactions  not only of  \emph {``mechanical"} type but also of thermodynamic type, that is, the \emph{``temperature effect"} due to the presence of different ``\emph{internal energy}". In the case in which all particles have the same temperature $u_i = u_\infty$,  the first two equations in \eqref{C-4} becomes the Cucker-Smale  model \eqref{C-0} with $\kappa = \frac{\kappa_1}{N u_\infty}$.

On the other hand, under the small diffusion velocity assumption $|\bv_i - \bv_c| \ll 1$ and by taking the barycenter observer such that  $\bv_c = 0$, we can further simplify the temperature equation in system \eqref{C-4} and derive an approximate model:
\begin{equation}\label{C-4-2}
\begin{cases}
\displaystyle \frac{d\bx_i}{dt} = \bv_i, \\
\displaystyle \frac{d\bv_i}{dt} = \frac{\kappa_1}{N} \sum_{j=1}^{N} \phi_1(\| \bx_j - \bx_i \|) \Big( \frac{\bv_j}{u_j} -  \frac{\bv_i}{u_i} \Big), \\
\displaystyle \frac{du_i}{dt} = \frac{\kappa_2}{N}  \sum_{j=1}^{N} \phi_2(\| \bx_j - \bx_i \|)
\Big( \frac{1}{u_i} - \frac{1}{u_j} \Big).
\end{cases}
\end{equation}

%%%%%%%%%%%%%%%%%%%%%%%%%%%%%%%%%%%%%%%%%%%%%%%%%%%%%%%%%%%%%%
\subsection{Thermodynamic Kuramoto model}\label{sect:3.3}
%%%%%%%%%%%%%%%%%%%%%%%%%%%%%%%%%%%%%%%%%%%%%%%%%%%%%%%%%%%%%%
In this subsection, we present the thermodynamic Kuramoto model \cite{[HPRS20]}  which can be derived from the two-dimensional thermodynamic Cucker-Smale model with constant communication weight and uniform modulus ratio $\frac{T_i}{u_i}$ between velocity and temperature:
 \[  \psi_{ij}:~\mbox{constant} \quad  \Big| \frac{\bv_i}{u_i} \Big| = \Big| \frac{\bv_j}{u_j} \Big|, \quad i,j = 1, \cdots, N. \]
 Now, we introduce a heading angle (or phase) $\theta_i$ such that
 \begin{equation} \label{C-4-3}
 \frac{\bv_i}{u_i} = \frac{\eta}{u_*} e^{{\mathrm i} \theta_i},  \quad \mbox{or equivalently} \quad \bv_i = \eta \frac{u_i}{u_*} e^{{\mathrm i} {\theta_i} },
 \end{equation}
 where $\eta$ is a constant parameter with speed dimension, and $u_*$ is a constant reference temperature.

We substitute the ansatz \eqref{C-4-3} into $\eqref{C-4-1}_2$ and compare the imaginary part of the resulting relation to get
\[ u_i \dot{\theta}_i = \frac{\kappa_1}{N}\sum_{j =1}^{N} \psi_{ij}\sin(\theta_j -\theta_i). \]
Once all temperatures $u_i$ are strictly positive, then the above relation can be rewritten as
\begin{equation} \label{C-4-4}
{\dot \theta}_i =  \frac{\kappa_1}{N}\sum_{j=1}^{N} \frac{\psi_{ij}}{u_i} \sin(\theta_j-\theta_i).
\end{equation}
Like the Kuramoto model \cite{[K75]}, we may add a natural frequency $\nu_i$ in the R.H.S of \eqref{C-4-4} to model the heterogeneity of particles and  obtain
\begin{equation} \label{C-4-5}
{\dot \theta}_i =  \nu_i + \frac{\kappa_1}{N}  \sum_{j=1}^{N} \frac{\psi_{ij}}{u_i}\sin(\theta_j -\theta_i).
\end{equation}
For the evolution of temperature $u_i$, we first use relation \eqref{C-4-3} to get
\begin{equation} \label{C-4-6}
\frac{1}{2} |\bv_i|^2 = \frac{\eta^2}{2u_*^2} |u_i|^2, \quad \mbox{i.e.,} \quad \frac{d}{dt}  |\bv_i|^2 = \frac{2\eta^2}{u_*^2} u_i {\dot u}_i.
\end{equation}
Then, we use $\eqref{C-4-1}_3$ and \eqref{C-4-6} to find the dynamics of $u_i$:
\begin{equation} \label{C-4-7}
{\dot u}_i = \frac{\kappa_2 u_*^2}{N(u_*^2+\eta^2u_i)}\sum_{j=1}^{N} \zeta_{ij}\left(\frac{1}{u_i}-\frac{1}{u_j}\right).
\end{equation}
Finally, we combine \eqref{C-4-5} and \eqref{C-4-7} to derive the thermodynamic Kuramoto model:
\begin{equation} \label{C-4-8}
\begin{cases}
\displaystyle \dot{\theta}_i = \nu_i +\displaystyle\frac{\kappa_1}{N}\sum_{j=1}^{N} \frac{\psi_{ij}}{u_i} \sin(\theta_j -\theta_i), \\
\displaystyle \dot{u}_i = \frac{\kappa_2}{N}\sum_{j=1}^{N} \frac{\zeta_{ij}u_*^2}{(u_*^2+\eta^2u_i)}\left(\frac{1}{u_i}-\frac{1}{u_j}\right),
\end{cases}
\end{equation}
where network topologies $(\psi_{ij})$ and $(\zeta_{ij})$ are assumed to be positive and symmetric:
\begin{equation} \label{B-9}
\psi_{ij}=\psi_{ji} > 0, \qquad \zeta_{ij}=\zeta_{ji} > 0, \quad 1 \leq i, j \leq N.
\end{equation}
We refer to \cite{[HPRS20]} for a detailed emergent dynamics of \eqref{C-4-8}.

%%%%%%%%%%%%%%%%%%%%%%%%%%%%%%%%%%%%%%%%%%%%%%%%%%%%%%%%%%%%%%
\subsection{Spinning Cucker-Smale particles}\label{sect:3.4}
%%%%%%%%%%%%%%%%%%%%%%%%%%%%%%%%%%%%%%%%%%%%%%%%%%%%%%%%%%%%%%

In this subsection, we present the Cucker-Smale model \cite{[HKKS19]} for the spinning particles. In this case, the internal variable $\bu_i$ is the spin of the $i$-th particle.
Then, the dynamics of the state $(\bx_i, \bv_i, \bu_i)$ is governed by the inertial spin model:
\begin{equation} \label{C-5}
\begin{cases}
\displaystyle \frac{d\bx_i}{dt} = \bv_i,  \\[3mm]
\displaystyle \frac{d\bv_i}{dt} = \frac{1}{\chi} \bu_i \times \bv_i, \\[3mm]
\displaystyle \frac{d\bu_i}{dt} = \bv_i \times \Big[ \frac{\kappa}{N} \sum_{j=1}^{N} p_{ij} (\bv_j - \bv_i)  - \gamma \frac{d\bv_i}{dt}  \Big],
\end{cases}
\end{equation}
where $\chi > 0$ is the generalized moment of inertia. Here, $p_{ij}$ represents a bounded communication weight or an interaction rate between $i$-th and $j$-th particles, and $\gamma$, and $\kappa$ represent strengths of damping, and coupling, respectively.  The friction term $-\gamma {\dot v}$ in spin dynamics  induces a rotational dissipation so that total spin tends to zero exponentially fast. Moreover, in a formal zero inertia limit $\chi \to 0$, system \eqref{C-5} with metric dependent weight $p_{ij} = \psi(\|\bx_i - \bx_j\|)$ reduces to the C-S type model with constant speed constraint \eqref{C-1-1}:
\begin{equation}
\begin{cases} \label{C-5-1}
\displaystyle \frac{d\bx_i}{dt} = \bv_i, \\[3mm]
\displaystyle \frac{d\bv_i}{dt} = \frac{{\bar \kappa}}{N}  \sum_{j=1}^{N} \psi(\| \bx_i - \bx_j \|) \Big(\bv_j - (\bv_j \cdot \bv_i) \bv_i \Big), \quad {\bar \kappa} :=  \frac{\kappa}{\gamma}.
\end{cases}
\end{equation}

%%%%%%%%%%%%%%%%%%%%%%%%%%%%%%%%%%%%%%%%%%%%%%%%%%%%%%%%%%%%%%%%
%%%%%%%%%%%%%%%%%%%%%%%%%%%%%%%%%%%%%%%%%%%%%%%%%%%%%%%%%%%%%%%%
\section{Beyond Cucker-Smale models: Case studies}\label{sect:4}
%%%%%%%%%%%%%%%%%%%%%%%%%%%%%%%%%%%%%%%%%%%%%%%%%%%%%%%%%%%%%%%%
%%%%%%%%%%%%%%%%%%%%%%%%%%%%%%%%%%%%%%%%%%%%%%%%%%%%%%%%%%%%%%%%

 This section shows how swarm  models can be derived referring to the mathematical structures presented in Section \ref{sect:2}. The presentation refers to specific case studies which are followed by numerical simulations. The model accounts for the influence over the motion of heterogeneously distributed activity variable corresponding to a specific social state. The activity variable is modified by the interaction dynamics. A plane dynamics is studied just to simplify notations,  however models can be rapidly transferred to three space dimensions.

 In more details, we  consider a  system constituted by one FS of $n$ interacting particles which heterogeneously share a social state.  This internal variable modifies the trajectories of their motion. The models is somehow inspired to human crowd dynamics as interactions account, at least partially,  for a dynamics which appears in crowds~\cite{[ALBI19],[ABGR20],[BGO19]}. The derivation accounts for the so-called topological interactions introduced in~\cite{[BCC08]}, see also~\cite{[BH17],[BS12]}.  The presentation  is delivered in the next subsections, where the first one deals with modeling, the second one  presents some simulations.

%%%%%%%%%%%%%%%%%%%%%%%%%%%%%%%%%%%%%%%%%%%%%%%%%%
\subsection{Modeling case studies}\label{subs:4.1}
%%%%%%%%%%%%%%%%%%%%%%%%%%%%%%%%%%%%%%%%%%%%%%%%%%

The derivation of models is here developed with the aim of showing, by a number of simple case studies, how the dynamics of the activity variable modifies the patterns of the flow dynamics. Specific models can be derived within the  following framework:

\begin{enumerate}

\item  The state of the system is given by the whole set  of directions $\theta_i$, rotation speeds $\sigma_i$ and activities $u_i$ of all a-particles with $i = 1, \ldots, n$, while each a-particle in the swarm moves with the same speed $v = v_0 = 1$. The activity $u_i$  is supposed to correspond to the level of stress with  $u_i \in [0,1]$, where  $u_i=0$ defines the lack of stress and  $u_i=1$  the maximal admissible level.

\vskip.2cm \item All a-particles  have a visibility angle, hence a visibility  domain $\Omega_i = [\theta_i - \Theta, \theta_i + \Theta]$ with radius $R$, each a-particle interacts only with other particles in $\Omega_i$ or even with a fixed ``small'' number of entities within a sensitivity domain $\Omega_i^s \subseteq \Omega_i$, where the radius $R^s$ is finite and depends of the number of particles selected for the interaction.

\vskip.2cm \item  Each a-particle has an individual movement  direction $\bnu_i$, but all of them  share a preferred  direction $\bnu$ which might depend of time. All particles have a trend towards  $\bnu$, but each of them is subject to an attraction towards the mean velocity direction $\bomega_i$ of the particles in the sensitivity domain. This attraction depends also on the social state of the particle which is heterogeneously distributed over the swarm and hence in $\Omega_i$.

\vskip.2cm \item The decision process by which a-particles modify their motion develops according to the following sequence: firstly the a-particle modifies the activity and subsequently the direction of motion.

\vskip.2cm \item The dynamics of the social state depends on  the specific features of the specific system under consideration. A simple case corresponds to a consensus dynamics with respect to the a-particles in the sensitivity domain.

\end{enumerate}

Let us now transfer these simple  rules into interactions models, to be inserted into the general structures proposed in Section \ref{sect:2}, referring specifically to Eq. \eqref{swarm-h} which needs some technical modifications (simplifications) to account for the assumptions in the above items.

The assumption that the same speed is shared by all particles defines the velocity of the i-particle as follows:
\begin{equation}
\bv_i = \cos \, \theta_i\, \bbi + \sin  \, \theta_i\, \bbj \hskip.5cm \Rightarrow \hskip.5cm
\frac{d\bv_{i}}{d t} = (- \sin \, \theta_i\, \bbi + \cos \, \theta_i\, \bbj) \, \sigma_i,
\end{equation}
where $\bbi$ and $\bbj$ denote the unit vectors of an orthogonal frame and $\theta_i \in [0, 2\pi)$ denotes the velocity direction. Hence the mechanical state of each particle is defined by flight direction $\theta_i$ and rotational speed $\sigma_i$, namely the time derivative of $\theta_i$. In addition, the following physical quantities can be defined:

\vskip.2cm \noindent The interaction rates $\eta_{ij} \cong \eta_0$ are supposed to be a constant quantity subsequently weighted by the distance between the interacting entities:
\begin{equation}\label{sens-decay}
 g = g(\bx_j, \bx_i;\alpha) = \exp\big\{- \alpha\, \big(||\bx_j - \bx_i||)\big\},
\end{equation}
where $\alpha$ is a positive defined parameter.

\vskip.2cm \noindent The attraction direction $\omega_i^s$ by the a-particles in $\Omega_i$ decays with the distance between the interacting particles:
\begin{equation}\label{sens-direction}
\omega_i^s  = \omega_i^s(\bx_i, \theta_i) = \sum_{j \in \Omega_i}\, g(\bx_j, \bx_i;\alpha) \, \theta_j.
\end{equation}

\vskip.2cm \noindent  The direction $\omega_i$, which effectively attracts the movement direction of the i-particle, is defined by a convex combination of $\theta^\nu$ and $\omega_i^s$
weighted by the activity
\begin{equation}\label{attraction}
\omega_i = \omega_i(\bx_i, \theta_i, u_i;\bnu)  = u_i\, \theta^\nu + (1 - u_i) \, \omega_i^s(\bx_i, \theta_i),
\end{equation}
where $\theta^\nu$ is the angle related to commonly preferred direction.

\vskip.2cm  \noindent  The dynamics by a consensus of the i-particle to the j-particles in $\Omega_i$  depending on a parameter $\beta$ as follows:
\begin{equation}\label{activity-consensus}
 \frac{du_i}{dt} = \beta\, \sum_{j \in \Omega_i}\, \exp\big\{- \alpha\, \big(||\bx_j - \bx_i||)\big\}\, (u_j -u_i).
  \end{equation}
When we go beyond the consensus dynamics, then different models of interaction can be considered keeping, however, the decay with distance as in~(\ref{activity-consensus}).

\subsubsection{\bf First order models}

\begin{figure}
\begin{center}
 \includegraphics[width=0.9\textwidth]{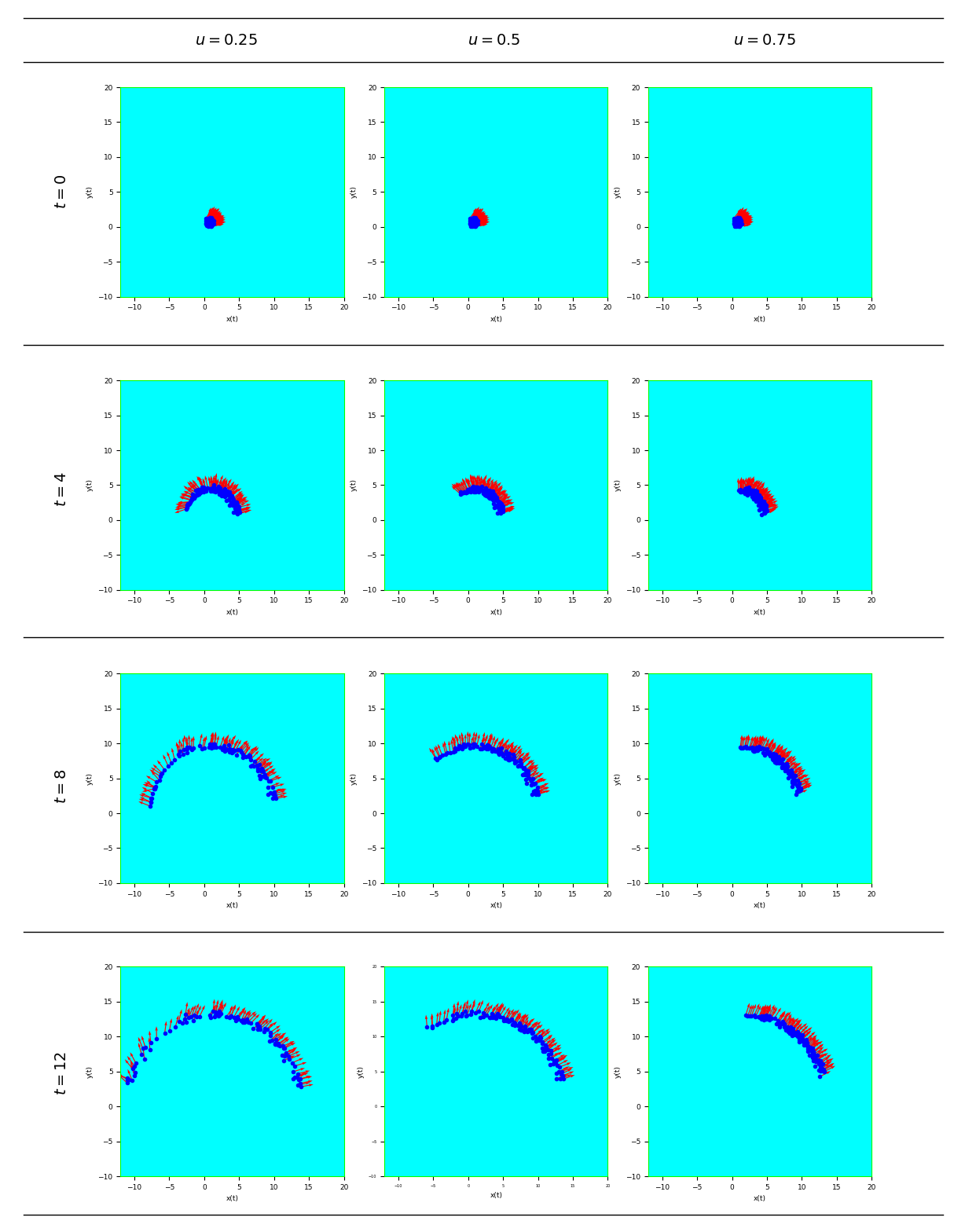}
\caption{Time dynamics of particles for different values of $u=0.25$, $u=0.5$ and $u=0.75$ corresponding to $t=0, 4, 8, 12$. Fixed preferred direction and constant activity.}
\label{fig:fig1}
\end{center}
\end{figure}

Let us now first consider  a first order model, where the alignment dynamic is modeled by a rotational speed depends, by a parameter $\gamma$, on the angular distance between the i-particle and that of the j-particles in $\Omega_i$:
\begin{equation}\label{alignment}
\frac{d\theta_i}{dt} = \gamma \, \big(\omega_i(\bx_i, \theta_i, u_i) - \theta_i \big)\pa
  \end{equation}

Let us specialize the components of $\bx$ by $\bx = \{x, y\}$ and  transfer the above assumptions into the mathematical structure (\ref{swarm-h}) yields:
\begin{equation}\label{swarm-A}
\begin{cases}
\displaystyle \frac{d u_i}{d t} = \beta \,\sum_{j \in \Omega_i}\, \exp\big\{- \alpha\, \big(||\bx_j - \bx_i||)\big\}\,(u_j -u_i),\\[3mm]
\displaystyle  \frac{dx_i}{dt} =  \cos \, \theta_i,\, \\[3mm]
\displaystyle  \frac{dy_i}{dt} =  \sin  \, \theta_i,\,  \\[3mm]
\displaystyle \frac{d\theta_i}{dt} = \gamma\, \big(u_i\, \theta^\nu + (1 - u_i) \, \sum_{j \in \Omega_i}\,\exp\big\{- \alpha\, \big(||\bx_j - \bx_i||)\big\} \, \theta_j - \theta_i\big).
 \end{cases}
\end{equation}

The following simplifications or generalizations can  be considered:

\vskip.2cm \noindent $\bullet$  {\bf The case of a constant activity:}  If $u_i$ attains the same value for all particles $u_i = u$ then the first equation has not an influence, while the system writes as follows:
\begin{equation}\label{swarm-B}
\begin{cases}
\displaystyle  \frac{dx_i}{dt} =  \cos \, \theta_i,\, \\[3mm]
\displaystyle  \frac{dy_i}{dt} =  \sin  \, \theta_i,\,  \\[3mm]
\displaystyle \frac{d\theta_i}{dt} = \gamma\, \big(u\, \theta^\nu + (1 - u) \, \sum_{j \in \Omega_i}\,\exp\big\{- \alpha\, \big(||\bx_j - \bx_i||)\big\} \, \theta_j - \theta_i\big),
 \end{cases}
\end{equation}
where, the very special case $u =1$ simplifies the alignment equation as follows:
\begin{equation}\label{alignment}
\frac{d\theta_i}{dt} = \gamma \, \big(\theta_i^\nu - \theta_i \big)\va
  \end{equation}
which is independent on the space variable. If $\ve = 1 - u$, where $\ve$ is small with respect to zero, then Eq.~(\ref{swarm-B}) can be treated as a perturbation, by the small parameter $\ve$, of the model corresponding to  $u =1$.

\vskip.2cm \noindent $\bullet$ {\bf Attraction to high values of $u$:} An additional possible modification of model \eqref{swarm-A} is that only high levels of $u_j$ are attractive, namely
\begin{equation*}
\begin{cases}
\frac{du_i}{dt} = \beta\, \sum_{j \in \Omega_i}\, \exp\big\{- \alpha\, \big(||\bx_j - \bx_i||)\big\}\, (u_j -u_i) &\text{ if } u_j > u_i \\[3mm]
\frac{du_i}{dt} = 0 &\text{ if } u_j \leq u_i.
\end{cases}
\end{equation*}
The model can be particularized in this case as
follows:
\begin{equation}\label{swarm-AA}
\begin{cases}
\displaystyle \frac{d u_i}{d t} = \beta \,\sum_{j \in \Omega_i}\, \exp\big\{- \alpha\, \big(||\bx_j - \bx_i||)\big\}\,(u_j -u_i)\delta_{\left\{u_j>u_i \right\}},\\[3mm]
\displaystyle  \frac{dx_i}{dt} =  \cos \, \theta_i,\, \\[3mm]
\displaystyle  \frac{dy_i}{dt} =  \sin  \, \theta_i,\,  \\[3mm]
\displaystyle \frac{d\theta_i}{dt} = \gamma\, \big(u_i\, \theta_i^\nu + (1 - u_i) \, \sum_{j \in \Omega_i}\,\exp\big\{- \alpha\, \big(||\bx_j - \bx_i||)\big\} \, \theta_j - \theta_i\big),
 \end{cases}
\end{equation}
where $\delta$ stand for the Kronecker delta symbol: $\delta_{\left\{u_j>u_i \right\}}=1$ if $ u_j > u_i$ and $\delta_{\left\{u_j>u_i \right\}}=0$ if $ u_j \leq  u_i$.

\subsubsection{\bf Second order models} Second order models can be developing by introducing a rotational speed $\sigma_i$ as the time derivative of $\theta_i$ and modeling the acceleration as follows:
\begin{equation}\label{alignment}
\frac{d\sigma_i}{dt} = \gamma_a \, \big(\omega_i(\bx_i, \theta_i, u_i) - \theta_i \big) - \gamma_b \frac{d\theta_i}{dt}\va
  \end{equation}
   corresponding to an acceleration induced by the chased and flight directions reduced by a viscous action, where $\gamma_a$ and $\gamma_b$ are a positive defined parameters. This phenomenological assumption yields:
\begin{equation}\label{swarm-C}
\begin{cases}
\displaystyle \frac{d u_i}{d t} = \beta \,\sum_{j \in \Omega_i}\, \exp\big\{- \alpha\, \big(||\bx_j - \bx_i||)\big\}\,(u_j -u_i),\\[3mm]
\displaystyle  \frac{dx_i}{dt} =  \cos \, \theta_i, \\[3mm]
\displaystyle  \frac{dy_i}{dt} =  \sin  \, \theta_i,  \\[3mm]
\displaystyle \frac{d\theta_i}{dt} = \sigma_i,  \\[3mm]
\displaystyle \frac{d\sigma_i}{dt} = \gamma_a \, \big(u_i\, \theta^\nu + (1 - u_i) \, \sum_{j \in \Omega_i}\,\exp\big\{- \alpha\, \big(||\bx_j - \bx_i||)\big\} \, \theta_j - \theta_i\big) - \gamma_b \frac{d\theta_i}{dt},
 \end{cases}
\end{equation}
while this model can be simplified under the assumption  of a constant activity.

\begin{figure}
\begin{center}
 \includegraphics[width=0.9\textwidth]{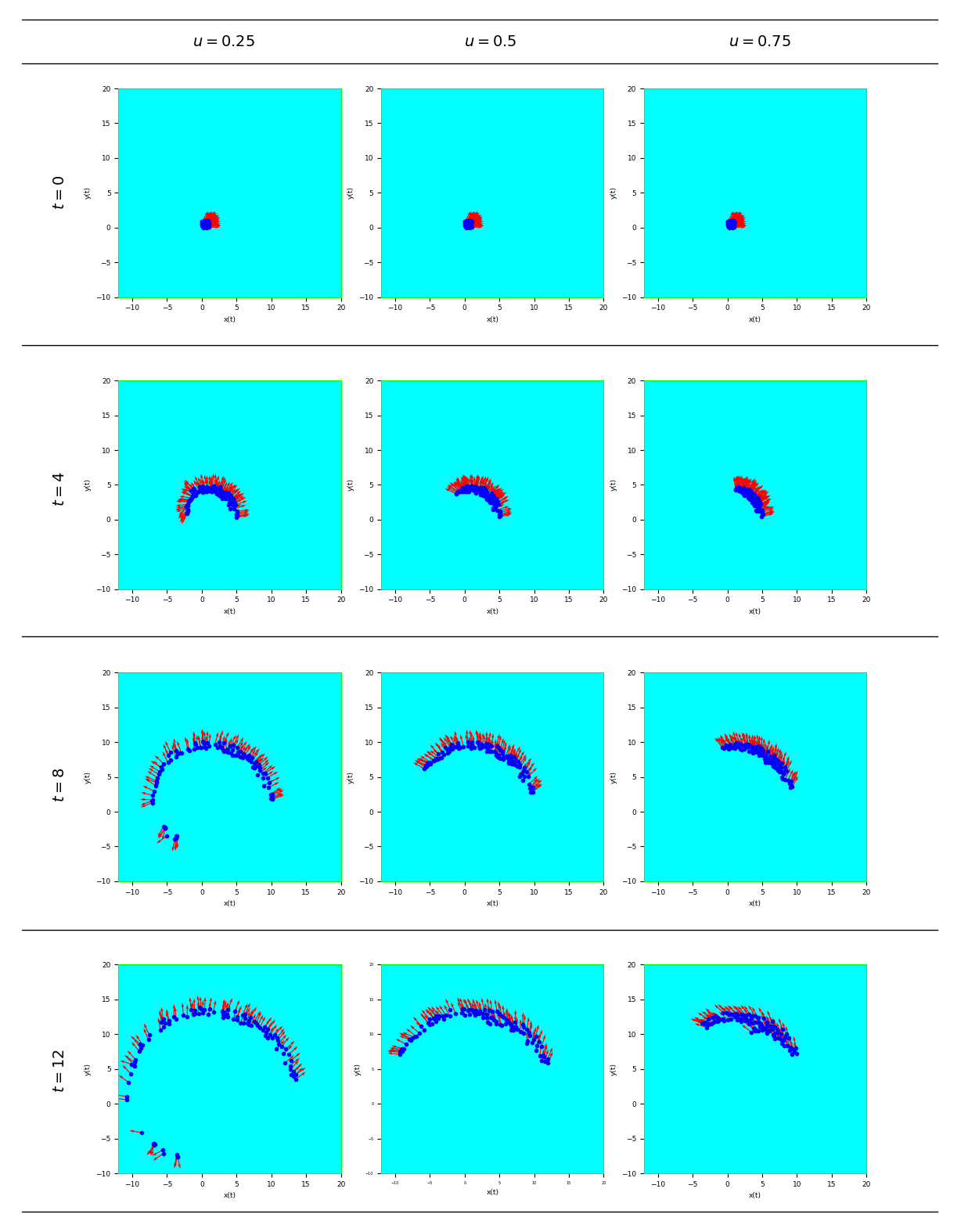}
\caption{Time dynamics of particles for different values of $u=0.25$, $u=0.5$ and $u=0.75$ corresponding to $t=0, 4, 8, 12$. Rotating preferred direction and constant activity.}
\label{fig:fig2}
\end{center}
\end{figure}

\subsubsection{\bf Topological interactions} Following~\cite{[BCC08]} interactions involve only a fixed number $m < n$ of i-particles within  the sensitivity domain $\Omega_i^s \subseteq \Omega_i$. The space dependence of the activity can be neglected in the modeling of interactions as  $m$ is of a smaller order with respect to $n$, do that $R^s$ is also small with respect to the visibility radius $R$.  In this case, the model simply writes as follows:
\begin{equation}\label{swarm-D}
\begin{cases}
\displaystyle \frac{d u_i}{d t} = \beta \,\sum_{j \in \Omega_i^s}\, (u_j -u_i),\\[3mm]
\displaystyle  \frac{dx_i}{dt} =  \cos \, \theta_i, \\[3mm]
\displaystyle  \frac{dy_i}{dt} =  \sin  \, \theta_i,  \\[3mm]
\displaystyle \frac{d\theta_i}{dt} = \gamma\, \big(u_i\, \theta^{\nu} + (1 - u_i) \, \sum_{j \in \Omega_i^s}\,(\theta_j - \theta_i\big) \big),
 \end{cases}
\end{equation}
while analogous calculations can be applied to the structures corresponding to the case of a constant activity.

%%%%%%%%%%%%%%%%%%%%%%%%%%%%%%%%%%%%%%%%
\subsection{Simulations}\label{subs:4.1}
%%%%%%%%%%%%%%%%%%%%%%%%%%%%%%%%%%%%%%%%
In this section, we provide a few numerical simulations on the  case studies treated in the preceding subsection, with the aim of showing how specific social dynamics modifies the patterns of the collective motion. More in detail, the focus is on the influence of the stress (activity) variable over the alignment of particles and on  how it contribute to the flocking behaviors.

\subsubsection{\bf On the role of constant activity} \label{sec:sec4.2.1}
 Let us refer to the first order model \eqref{swarm-B}, sample simulations are carried out which aim at enlightening how the activity variable, which act as a parameter, as it has a constant value $u$, can affect the collective dynamics. Two specific cases are presented in the following.

\vskip.5cm \noindent $\bullet$ {\bf Fixed preferred  direction:} Simulations are developed
corresponding to $n=100$ a-particles and to a fixed values of the preferred  direction, specifically the value $\theta^\nu=\frac{\pi}{3}$ is selected. Initially, the active particles are randomly distributed in a square region of dimension $[0,1] \times [0,1]$ while the velocity direction of each a-particle have a random distribution in $\Omega = \left[\theta^\nu - \frac{\pi}{3}, \theta^\nu + \frac{\pi}{3}\right]$. The parameters used for the simulations are $\alpha=1$, $ \gamma=0.1$ and $\Theta=\frac{\pi}{4}$.

Simulations are shown in Figure \ref{fig:fig1}  which reports the flow patterns for $u=0.25$, $0.5$, $0.75$ corresponding to $t=0$, $4$, $8$, $12$. It can be observed that high values of $u$ contribute to flocking and alignment toward the shared preferred direction, while decreasing values of $u$ reduce the alignment of the active particles.

\vskip.5cm \noindent $\bullet$ {\bf Rotating preferred  direction:}
Consider the case of a rotating direction of the  preferred velocity. The study is analogous to that treated above, but with the only difference that now the preferred direction shared by all particles rotates anticlockwise with constant angular velocity. Therefore, the  model is the same given in Eq. \eqref{swarm-B}  with $\bnu (t)= \left(\cos (a_0t),\sin (a_0t)\right)$, where $a_0$ is the angular velocity, hence the angle related to commonly preferred direction is given by:
\begin{equation*}
\theta^\nu(t)=a_0t.
\end{equation*}

Similarly to the case of a constant velocity direction simulations refer to three different values of $u=0.25$, $0.5$, $0.75$ for $n=100$ active particles which at $t=0$ are  randomly distributed in the unit square. The initial velocity direction has a random distribution in $\Omega$. Other parameters are  $\alpha=1$, $ \gamma=0.1$, $\Theta=\frac{\pi}{4}$ and $a_0=0.02$.

Simulations are shown in Figure \ref{fig:fig2}. Also in this case,  simulations show how different values of the activity can have an influence on the alignment dynamics of the active particles. In fact, it is apparent that decreasing values of $u$ reduce the alignment of the particles, while  high values of $u$ contribute to flocking. However, active particles appear to be rotating  in anticlockwise direction, and increasing values of $u$ significantly affect the rotation of the particles and  increase the radius of there  circular trajectory.

\subsubsection{\bf On the role of variable activity}

\begin{figure}
\begin{center}
 \includegraphics[width=\textwidth]{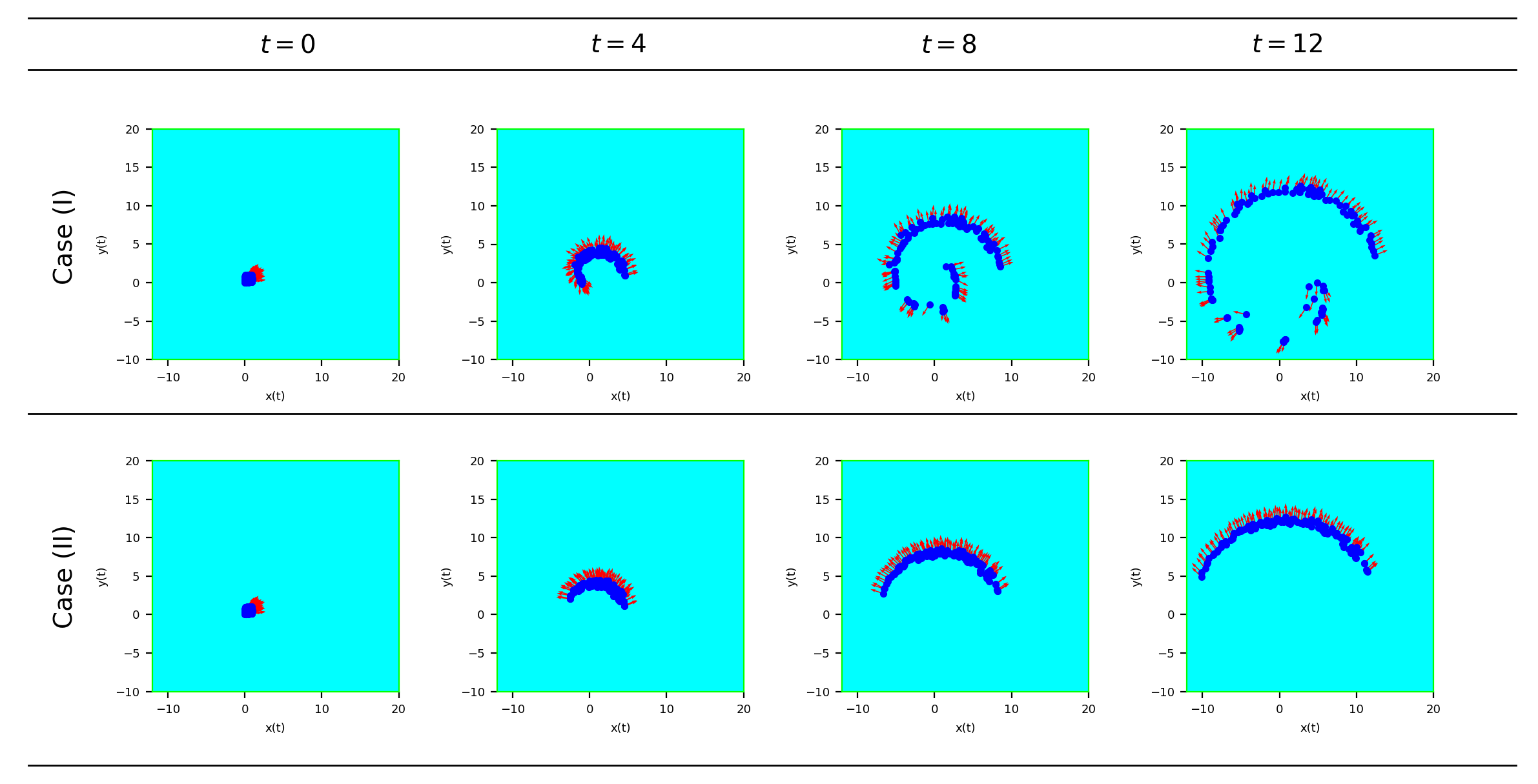}
\caption{Time dynamics of particles for different  values of $t=0, 4, 8, 12$. Fixed preferred direction and variable activity.}
\label{fig:fig3}
\end{center}
\end{figure}

We return now to model \eqref{swarm-A} where, compared to  \eqref{swarm-B}, the dynamics of each $u_i$ is now delivered by a consensus towards a mean activity, see Eq. \eqref{activity-consensus}.

Two cases are studied:
\begin{itemize}
\item[] Case (I): each $u_i$ is delivered by a consensus towards a mean activity (model \eqref{swarm-A}).\vspace*{0.2cm}
\vskip.2cm \item[]Case (II): attraction to high values of activity variable (model \eqref{swarm-AA}).\vspace*{0.2cm}
\end{itemize}

We consider initial data for the activity variable  randomly distributed in the interval $[0,1]$ and we chose  $\beta=0.1$. The other parameters are the same as in the case of a fixed preferred direction and a constant activity. Figure \ref{fig:fig3} reports the time dynamic of active particles for  $t=0$, $4$, $8$, $12$ in cases (I) and (II). In both cases a flocking behavior is observed and is more relevant in the second case, case (II), which clearly shows that attraction to high values of $u$ enhances flocking.

%%%%%%%%%%%%%%%%%%%%%%%%%%%%%%%%%%%%%%%%%%%%%
%%%%%%%%%%%%%%%%%%%%%%%%%%%%%%%%%%%%%%%%%%%%%
\section{Research perspectives}\label{sect:5}
%%%%%%%%%%%%%%%%%%%%%%%%%%%%%%%%%%%%%%%%%%%%%
%%%%%%%%%%%%%%%%%%%%%%%%%%%%%%%%%%%%%%%%%%%%%

A new class of swarm models has been derived in this paper. The main feature is that models where the micro-state of the interacting entities includes, in addition to mechanical variables - typically position and velocity - also additional variables which model the heterogeneous behavior by which the said entities develop their overall dynamics. These variables, in analogy with the kinetic theory of active particles~\cite{[BBGO17]}, can be called \textit{activity} which can be either a scalar or even a vector. Firstly the mathematical structures underlying the aforementioned behavioral dynamics have been derived and subsequently some sample models have been presented. Simulations have enlightened how, even in very simple cases, the dynamics shows large deviations corresponding to large deviations of the activity variables.

The rationale is analogous to that of the kinetic theory approach~\cite{[BBGO17]} which requires a large number of interacting active particles to justify the derivation of models. On the other hand, the derivation of swarm models does not require this specific assumptions. This approach has already generated some models of interest in economy concerning the role of idiosyncratic learning of firms as an engine for the derivation of market sharing~\cite{[BDKV20]} or the modeling of price dynamics in the complex interaction between buyers and sellers~\cite{[BDKMT19]}.

Looking at research perspectives we can identify three key problems which appear to us worth to be chased in a possible research quest.

\begin{itemize}

\item \textit{Derivation of a more general framework suitable to include the main key features of living systems.}

\vskip.2cm \item \textit{Applications to the modeling specific systems such as crowds in crisis situations, for instance evacuation under stress conditions~\cite{[BGO19]} or under contagion risk~\cite{[BBC20],[KQ19]}}.

\vskip.2cm \item \textit{Study of analytic problems within a multiscale vision where, by suitable asymptotic time-space dynamics and continuity assumptions, firstly kinetic type models are derived from the underlying description and subsequently hydrodynamical models are derived from the underlying kinetic description~\cite{[BC19]}.}

\end{itemize}

It is plain that these key problems have been generated by the authors bias which looks ahead to the perspectives to develop the approach introduced in our paper toward a mathematical theory of dynamical systems which aims at modeling the complex behavior of living systems.

\end{document}